\documentclass[]{article}
\usepackage{lmodern}
\usepackage{amssymb,amsmath}
\usepackage{ifxetex,ifluatex}
\usepackage{fixltx2e} 
\ifnum 0\ifxetex 1\fi\ifluatex 1\fi=0 
  \usepackage[T1]{fontenc}
  \usepackage[utf8]{inputenc}
\else 
  \ifxetex
    \usepackage{mathspec}
  \else
    \usepackage{fontspec}
  \fi
  \defaultfontfeatures{Ligatures=TeX,Scale=MatchLowercase}
\fi
\IfFileExists{upquote.sty}{\usepackage{upquote}}{}
\IfFileExists{microtype.sty}{%
\usepackage{microtype}
\UseMicrotypeSet[protrusion]{basicmath} 
}{}
\usepackage[margin=1in]{geometry}
\usepackage{hyperref}
\hypersetup{unicode=true,
            pdftitle={The Impacts of the Alaska Permanent Fund Dividend on High School Status Completion Rates},
            pdfauthor={Mattathias Lerner},
            pdfborder={0 0 0},
            breaklinks=true}
\usepackage{longtable,booktabs}
\usepackage{graphicx,grffile}
\makeatletter
\def\maxwidth{\ifdim\Gin@nat@width>\linewidth\linewidth\else\Gin@nat@width\fi}
\def\maxheight{\ifdim\Gin@nat@height>\textheight\textheight\else\Gin@nat@height\fi}
\makeatother
\setkeys{Gin}{width=\maxwidth,height=\maxheight,keepaspectratio}
\ifx\paragraph\undefined\else
\let\oldparagraph\paragraph
\renewcommand{\paragraph}[1]{\oldparagraph{#1}\mbox{}}
\fi
\ifx\subparagraph\undefined\else
\let\oldsubparagraph\subparagraph
\renewcommand{\subparagraph}[1]{\oldsubparagraph{#1}\mbox{}}
\fi

\let\rmarkdownfootnote\footnote%
\def\footnote{\protect\rmarkdownfootnote}

\usepackage{titling}

\newcommand{\subtitle}[1]{
  \posttitle{
    \begin{center}\large#1\end{center}
    }
}

  \title{The Impacts of the Alaska Permanent Fund Dividend on High School Status
Completion Rates}
    \pretitle{\vspace{\droptitle}\centering\huge}
  \posttitle{\par}
  \subtitle{A Synthetic Control Study}
  \author{Mattathias Lerner}
    \preauthor{\centering\large\emph}
  \postauthor{\par}
      \predate{\centering\large\emph}
  \postdate{\par}
    \date{May 2019}

\usepackage{setspace}

\begin{document}
\maketitle

\begin{abstract}
\noindent Direct cash transfer programs have shown success as poverty interventions in both the developing and developed world, yet little research exists examining the society-wide outcomes of an unconditional cash transfer program disbursed without means-testing. This paper attempts to determine the impact of direct cash transfers on educational outcomes in a developed society by investigating the impacts of the Alaska Permanent Fund Dividend, which was launched in 1982 and continues to be disbursed on an annual basis to every Alaskan. A synthetic control model is deployed to examine the path of educational attainment among Alaskans between 1977 and 1991 in order to determine if high school status completion rates after the launch of the dividend diverge from the synthetic in a manner suggestive of a treatment effect.
\end{abstract}

\newpage

\section{Introduction}\label{introduction}

In 1976, the people of Alaska voted to amend their state's constitution
in order to allow the creation of a dedicated fund that would be managed
independently from the day-to-day spending of the government. The
purpose of the fund was to hold and invest the oil revenues that would
begin filling the state's coffers with the completion of the Trans
Alaska Pipeline the following year. The vote enshrined Article IX,
Section 15 of the Alaska Constitution, which holds that a minimum of
twenty-five percent of Alaska's mineral revenue are to flow directly
into the Permanent Fund.

In 1980, the Alaskan legislature created the Alaska Permanent Fund
Corporation, a quasi-independent entity tasked with managing the growing
assets of the Fund. The same year, the legislature established the
Permanent Fund Dividend (PFD), which pays a cash sum based on a
five-year average of the Fund's performance to all state residents who
have lived in Alaska for at least a full calendar year. Its first year,
the fund paid out \$1000 USD to all eligible residents---a sum roughly
equal to \$2600 today. The Alaska Permanent Fund is now valued at over
\$60 billion.

Dividends from the Fund are issued annually to every man, woman and
child in the state of Alaska, and have ranged over the course of its
existence from approximately \$300 to approximately \$2000. Since
petroleum production is the source of the fund's value, dividend payouts
vary with the price of oil. In addition, the State has some latitude to
divert money from the fund and to influence the size of the dividend. In
2008, Alaska's state legislature approved an additional one-time payment
of \$1200 from the fund to all the state's residents.

In-depth empirical investigation of the Alaska Permanent Fund holds
substantial interest for contemporary political discourse. Though the
size of the dividend has historically been too small to constitute a
living wage, mention of the Fund has begun to appear in recent
conversations about the concept of universal basic income and, more
broadly, in discussions about the role of cash transfers in shaping
development outcomes. Modest basic income trials have recently been
initiated in several locales (e.g.~Goodman, 2018). The ideological
versatility of these programs has been widely lauded: American
libertarians view cash transfers as a salve for the inefficiencies of
the bureaucratic welfare state (Zwolinski, 2014) and democratic
socialists see social wealth funds (a scenario in which dividends would
come from funds built through government purchases of ownership in
private companies using tax revenue from high earners) as a pragmatic
path to public ownership of the means of production (Bruenig, 2017).

Some previous research into the virtues of Alaska's dividend has focused
on labor market outcomes: since a common criticism of unconditional cash
transfers is that they reduce incentives to work, economists have sought
relationships between changes in part- and full-time employment and the
disbursement of the dividend. An NBER working paper from this year
(Jones \& Marinescu, 2018) found no significant reduction in aggregate
employment as a result of the dividend. Another recent study (Berman,
2018) identified significant reductions in the poverty rate among
Alaska's Native population.

There is considerable research on the impact of cash transfers on a wide
variety of vital measures in developed and developing economies, from
birth outcomes (Amarante, Manacorda, Miguel \& Vigorito, 2011) and
education (Edmonds, 2006) to entrepreneurship (de Mel, McKenzie \&
Woodruff, 2007) and mental health (Baird, Feirreira, Özler \& Woolcock,
2013). In North America, this research is augmented by a large body of
studies investigating the impact of the Earned Income Tax Credit (EITC),
the results of trials designed to assess the impacts of a negative
income tax (NIT), and analysis of data produced by the Manitoba Basic
Annual Income Experiment (``Mincome'').

In this paper, I investigate one avenue by which the disbursement of the
PFD could potentially influence social welfare: educational attainment.
As an outcome variable, I use high school status completion rate, which
I calculate using the Current Population Survey for each available state
and year in the timeframe 1977-1991. The high school status completion
rate is defined as the proportion of 18- to 24-year-olds who hold a high
school diploma or alternative credential. Using a synthetic control
model, I attempt to discern differences between educational attainment
trends in Alaska and in a ``synthetic Alaska''---a counterfactual
version of the state in which no dividend is disbursed.

\section{Literature Review}\label{literature-review}

\subsection{Earned Income Tax Credit}\label{earned-income-tax-credit}

Though no directly comparable program to the Alaska Permanent Fund
Dividend exists elsewhere in the United States, the Earned Income Tax
Credit (EITC) arguably provides a strong prior basis for believing that
the PFD might result in positive social welfare impacts. The EITC is a
federal tax credit, begun in 1975, and is considered by some to be the
``cornerstone of U.S. anti-poverty policy'' (Hoynes 2014) and the ``most
effective anti-poverty program for working families'' (Eckholm 2007).
This latter claim is important: the EITC is effectively an earnings
subsidy, and individuals who do not work are not eligible to receive the
transfer. The EITC initially came about, in part, as a reaction to
proposals for a negative income tax (NIT), a policy most forcefully
advocated for by Milton Friedman (Moffitt 2003). Though NIT payments
would have reduced with increased work hours, early EITC proponents
believed that the main benefits of the NIT would accrue to those without
employment, and would therefore reduce work incentives (Hotz \& Scholz
2003); the EITC proved to be a more politically palatable alternative.

In its four decades of existence, the EITC has been seen by many to have
been a resounding success. In 1997 and 1998, for example, the program
lifted more than 4 million families out of poverty (Council of Economic
Advisers 1998). Perhaps more compellingly for the current research, a
marginal \$1000 EITC exposure for children between 13 and 18 is
associated with increased odds of completing high school and college
(Bastian \& Michelmore 2018). This is congruent with other research
suggesting that EITC exposure in a student's senior year of high school
is associated with an increase in college enrollment the following year
(Manoli \& Turner 2018), and that the credit seems to improve elementary
school students' test scores (Chetty, Saez \& Rockoff 2011). A larger
body of research, though outside the scope of the present study,
connects the EITC to a variety of other positive welfare outcomes, such
as a decreased number of low-birthweight births (Hoynes, Miller \&
Simon, 2017).

Manoli \& Turner (2018) propose that credit constraints and excess
sensitivity to cash-on-hand partially explain the apparently large
effect of EITC payments on educational attainment in some settings. They
observe that university enrollment involves up-front, out-of-pocket
costs that may not be anticipated or planned for by families in
financially constrained circumstances, and note that this observation is
consistent with the finding that the apparent income effects on college
enrollment do not persist at higher income levels. Support for an
analogous circumstance in the case of high school completion can be
found in the educational literature, such as the work of Doll, Eslami \&
Walters (2013), who conduct an exhaustive review of research on high
school non-completion. These authors note that that students who drop
out toward the end of their high school educations often cite the
incompatibility of work and school as their reason for dropping out. If
the EITC and other comparable programs provide sufficient financial
cushion in order to bring work and school out of conflict, an increase
in completion rates could result.

\subsection{NIT Trials}\label{nit-trials}

The EITC as it currently exists and the NIT as it was proposed are
subtly different programs. In its simplest initial formulations by
Milton Friedman (Friedman 2002; Friedman 1987), the NIT would have
entailed a fixed amount of allowances---essentially a minimum income
level---and a subsidy rate governing how much of the gap between a
family's earned income and the minimium income level would be made up
with government assistance. For example, with an allowance of \$20,000
and a subsidy rate of 20\%, a family earning \$10,000 would receive
\$2,000 from the government; a family earning \$15,000 would receive
\$1,000 and a family earning \$19,000 would receive \$200.

Friedman saw this setup as a potential replacement for a bricolage of
federal welfare programs that seemed to create situations in which
working more resulted in overall benefit reductions for poor families.
Yet the simplicity of a single, federally mandated minimum income level
may have been the death of the idea: income levels high enough to
protect urban working families would have been enough to comfortably (if
not luxuriously) support rural loafers, creating a politically
unsustainable state of affairs (Frank 2006).

During the heyday of the NIT debate in the United States, several
small-scale experiments were conducted to investigate the impact of such
a program on social welfare. One such program was conducted in Gary,
Indiana, where positive effects on school performance were found for
fourth- through seventh-graders, with increased effects associated with
longer time spent in the program (Maynard \& Murnane, 1979). Another
study, known as the Seattle-Denver Income Maintenance Experiment
(SIME/DIME), found increases in school attendance associated with
participation in a similar minimum income program (Venti, 1984). The New
Jersey Experiment in Income Maintenance, conducted around the same time,
found increases in educational attainment associated with program
participation (Hanushek, 2003). Hanushek (2003) notes that one possible
cause of increased school attendance may be that income subsidies reduce
the opportunity cost of attending school.

The findings of Venti (1984), in particular, lend credence to the
suggestion, made above, that income support can lead to a substitution
of school for work in high school students; Venti finds strong and
statistically significant reductions in the probability of working for
16- to 21-year-olds that are offset by an increase in school attendance.
In contrast, Maynard \& Murnane (1979) find no effect on older children,
but they stress that the samples under review---in Gary, Indiana, and in
rural North Carolina---are not representative of the U.S. student
population as a whole.

The transferability of results from NIT trials to the PFD case is
promising but unclear, since the transfers are of qualitatively
different kinds. The NIT, like the EITC, is not universal; indeed, few
cash transfer programs are. More importantly, however, the NIT
experiments were explicitly short-term experiments, and it is unclear
whether families responded differently in that knowledge. Would these
families have smoothed their consumption differently if the trial had
instead been a permanent program? The answer to this question is
impossible to know, but bears directly on the generalizability of these
findings to the Alaskan case.

\subsection{Non-EITC Cash Transfer
Programs}\label{non-eitc-cash-transfer-programs}

A growing body of research, largely in the developing world but
including some more recent work in developed countries, provides
evidence for various positive welfare impacts of unconditional cash
transfer programs. In Uruguay, researchers found that a series of
transfers amounting to roughly a doubling of monthly income was
associated with a 1.7\% reduction in low-birthweight births among
mothers who received the transfer (Amarante, Manacorda, Miguel \&
Vigorito, 2011), and similar studies elsewhere have found that cash
transfers can positively influence structural determinants of overall
health (Owusu-Addo, Renzaho \& Smith, 2018). However, meta-analyses have
found inconclusive evidence of summary health effects of cash transfer
programs, at least in lower-middle-income countries (Pega et al., 2017).

With respect to education, some evidence from the developing world
points to substantial positive impacts on attendance and enrollment from
unconditional transfer programs (Benhassine, Devoto, Duflo, Dupas \&
Pouliquen, 2013), while other research indicates limited enrollment
effects and slight positive impacts on completion rates (Araujo, Bosch
\& Schady, 2016).

Unconditional cash transfer programs have also been attempted on a
preliminary basis in the developed world. A recent basic income trial in
Finland yields some positive social welfare impacts (for example, on
well-being) but no positive conclusions regarding education (Finnish
Ministry of Social Affairs and Health, 2019). In the United States, the
Eastern Band of Cherokee Indians, located in North Carolina, has
distributed a portion of casino profits to all adult tribal members
since 1997; these disbursements are associated with a significant
positive impact on educational attainment on the basis of an additional
income (on average) of \$4000 per year. Akee, Copeland, Keeler, Angold
\& Costello (2010) find that this additional income increases
educational attainment by 1 year at age 21.

One of the most well-known early experiments with unconditional cash
transfers in the developed world was conducted in Manitoba between 1974
and 1979. This program is more closely comparable to the Alaskan case
because the experiment included a saturation site---the town of
Dauphin---in which all eligible residents were entitled to receive a
cash transfer; nevertheless, only a third of residents qualified. This
experiment seems to have resulted in modestly higher educational
attainment among treated households, with an increase in the proportion
of students progressing from Grade 11 to Grade 12 (Forget, 2011);
research also indicated a decrease in hospitalizations, particularly
those related to mental health and accidental injury.

Forget (2011) notes a puzzle in the Mincome data that has direct
applicability to the Alaska case: though only a third of families in
Dauphin (the saturation site) qualified for payments, and though many of
these payments were small, an education effect is apparent and
statistically significant. Forget observes that the students most likely
to leave school came from the same low-income families that were
eligible for the Mincome stipend; high-income students were already
relatively likely to finish school. The same logic applies to the Alaska
case: though all families receive the dividend, an effect an education
could still be perceptible due to the payment's differential effect on
low-income students, e.g.~those most likely not to finish high school.

\subsection{Previous research on the
PFD}\label{previous-research-on-the-pfd}

In the three-and-a-half decades of the Alaska Permanent Fund Dividend,
researchers have investigated various aspects of the program. Two papers
in particular can be considered important precedents for my own: the
first is a well-known NBER working paper by Damon Jones and Ioana
Marinescu (2018), who fit a synthetic control model (upon which I
closely model my own) in order to estimate the effects of the PFD on
part- and full-time employment in Alaska. The second, by Matthew Berman
(2018), finds that the PFD has had a significant positive impact on
poverty among indigenous people in Alaska.

Jones and Marinescu use data from the Current Population Survey (CPS) to
form a synthetic control for Alaska from 1977 (the earliest year for
which CPS data is available specifically for Alaska) through 2014. They
find no difference in employment, labor force participation or hours
worked as a result of the PFD; however, they do find a statistically
significant increase in the part-time rate. In particular, the apparent
null effect on hours worked suggests that the hypothesized causal
pathway in which the PFD reduces work hours for high school students,
thereby bringing work and school out of conflict and increasing status
completion rates, may not be supported by the evidence.

Berman (2018) uses Census data to identify the portion of individuals'
income constituted by the PFD and compares it to their incomes without
the PFD; in this manner, he identifies the portion of indigenous
Alaskans moved above the poverty threshold by means of the dividend: for
some time periods, Berman finds differences as large as 10 percentage
points between Native poverty rates with and without the PFD.

Hsieh (2003) uses the PFD as a means to examine the permanent income
hypothesis, which holds that consumers should not respond to previously
anticipated changes in their income. Using data from the Consumer
Expenditure Survey (CEX), Hsieh finds evidence that Alaskan households
smooth their spending in a way that accords with the predictions of the
life-cycle/permanent income hypothesis. This finding is accompanied by
evidence that the same sample of houses do \emph{not} appear to smooth
their incomes in response to expected income tax refunds. Hsieh
attributes this seeming discrepancy to bounded rationality, citing the
differing cognitive demands placed on households when they consider tax
refunds and the PFD, respectively. The Permanent Fund Dividend is
distributed annually and without fail; moreover, since its amounts are
based on a five-year average of oil revenues, year-to-year amounts are
not extremely volatile and are therefore easily foreseen even by
unsophisticated individuals.

Hsieh's work is therefore in some tension with the mechanism proposed
for the educational impact of the EITC by Manoli \& Turner (2013). If
Alaskan households successfully smooth their consumption in anticipation
of receipt of the PFD, then the dividend funds are not available as
excess cash-on-hand in event of unanticipated expenses at the time of
college enrollment, or at any other time. More broadly, if the permanent
income hypothesis holds with respect to the PFD, then the availability
of excess cash-on-hand is unlikely to be the potential mechanism for any
kind of social welfare impact of the PFD.

This preceding work on the PFD is important in the present context since
it helps to provide a basis for believing---along with aforementioned
data on the EITC and NIT experiments---that the relatively modest size
of the dividend can have statistically perceptible effects at a large
scale. At the same time, Hsieh (2003) provides reason for believing that
the mechanisms that might lead to a positive education effect for the
EITC might not hold for the PFD.

\section{Methodology}\label{methodology}

In this paper, I work largely along the lines of Jones and Marinescu,
deploying the synthetic control model popularized Abadie and Gardeazabal
(2003) and extended by Abadie, Diamond \& Hainmueller (2010). For
mathematical details on the method, readers should refer to those
papers. In this paper, I use the R package \emph{Synth} (Abadie, Diamond
\& Hainmueller, 2011).

My choice of methodology here results from the challenges of causal
inference in the PFD scenario. The variable under investigation is a
state-level statistic (youth educational attainment), and we have only
one treated unit (the state of Alaska). For this reason, a conventional
difference-in-differences approach is difficult to implement---and, in
any case, the required parallel trends assumption is difficult to
defend: state-level policies that affect educational attainment and
enrollment are continually being instituted, extended, modified, amended
and occasionally rolled back throughout the country; there is no reason
to accept the strong assumption that parallel trends in the pretreatment
would, even in the absence of the specified treatment, continue to hold
in control units in the post-treatment period. The synthetic control
method is more robust to this concern, for reasons elaborated on below.

The synthetic control method is a quasi-experimental technique used to
determine the effects, over time, of a policy implemented at a single
point in time on an experimental unit for which there is no suitable
real-life control. A ``synthetic control'' is constructed using a data
from a weighted pool of donor states. A broad outline of the method
follows below.

In the synthetic control method, the treated unit is compared to a
single synthetic control unit, which is constructed using a weighted
average of \(\it{m}\) candidate donor units. The goal is to determine
the effect of a treatment administered at time \(\it{t}\), using data
from \(\it{k}\) pretreatment periods and \(\it{q}\) post-treatment
periods, on some variable of interest \(\bf{Y}\). For a given outcome
metric, a set of \(\it{p}\) prognostic covariates \(\bf{V}\) is
selected; conventionally, this set includes lagged values of the
variable of interest, which are naturally highly predictive of current
values of a given time-varying measurement. An \(\it{(m \times 1)}\)
vector of weights \(\bf{W}\) is selected to minimize
\(\bf{||X_1 - WX_0||}\), where \(\bf{X_1}\) is a \(\it{(p \times 1)}\)
vector in which each \(X_{1_n}\) is some linear combination of the
values of covariates \(\bf{V}\) over the pretreatment period
\(\it{(t - k, t)}\); \(\bf{X_0}\) is a \(\it{(p \times m)}\) matrix of
covariates for the pool of candidate regions with the same makeup. A
separate set of predictor weights is selected in order to minimize the
mean squared prediction error between the weighted predictors and the
values of \(\bf{Y}\) over the pretreatment period.

After goodness-of-fit testing, a placebo test is conducted on states in
the donor pool in order to determine the significance of the results
calculated for the region under investigation. Separate synthetic
controls are fit for all possible counterfactual treatment regions; for
each model, the root mean squared error (RMSE) is recorded and the
observed RMSE of the original model is compared with the distribution of
counterfactual RMSEs. In this way, a ``p-value'' is recorded that
suggests the likelihood of observing a difference as large as the
observed treatment-synthetic gap. A null hypothesis of no effect is not
tested; rather, a large difference is treated as being ``suggestive'' of
a treatment effect (McClelland and Gault 2017).

\section{Data}\label{data}

\subsection{Data Description}\label{data-description}

This study uses two main data sources. The first is an extract from the
Current Population Survey (CPS) for the years 1977 to 1991. The CPS is a
monthly survey of the civilian non-institutionalized population
conducted by the U.S. Bureau of Census on behalf of the the U.S. Bureau
of Labor Statistics. It collects a wide variety of household- and
individual-level variables relevant for statistical evaluation of the
U.S. labor force. 1977 is the earliest year available from CPS for the
purposes of this research: although a variable identifying a CPS
respondent's state of residence is available going back to 1962, CPS did
not disaggregate Alaska between 1968 and 1976 (inclusive); for those
years, it was included as an component of multi-state groupings
including, in the first case, Alaska, Washington, and Hawaii (1968-1972)
and, in the second case, Washington, Oregon, Hawaii, and Alaska
(1973-1976). Since Alaska cannot be meaningfully disaggregated for this
period, it is not available for use in the synthetic control model
deployed here.

The timeframe under investigation for this study ends in 1991. In 1992,
CPS switched its method for recording educational attainment. Though
recent CPS publications offer a recoded variable that incorporates both
pre- and post-1992 measurements, this recode does not allow for
comparability across time periods that include the 1992 measurement
change.

This study also uses the Correlates of State Policy dataset produced by
Michigan State University. The main utility of this data in the present
study is in providing predictors for the outcome variables for which I
create a synthetic control. The dataset aggregates state-level
indicators and metrics from a variety of sources for the years 1900 to
2016. However, few variables are available for the entire range of the
study (1977-1991), so only relatively modest use is made of this
dataset.

\subsection{Variable of interest}\label{variable-of-interest}

The time period under investigation offers some challenges in terms of
finding a suitable outcome variable due to measurement changes in the
United States over the past several decades. The National Center for
Education Statistics uses several similar but distinct metrics for
analyzing high school completion. At present, these are the event
dropout rate, the status dropout rate, the status completion rate, the
adjusted cohort graduation rate (ACGR), and the averaged freshman
graduation rate (AFGR). More recent iterations of the CPS have included
questions about present school enrollment or attendance; these questions
are not available prior to 1992, when a redesign of the CPS introduced a
new question intended to enable better calculation of the various
educational statistics listed above.

The status completion rate is here defined as the percentage of young
adults between the ages of 18 and 24 who hold a high school credential.
The primary weakness of this metric is that it makes no distinction
between individuals who were born and/or raised in Alaska and those who
migrated to the state. This weakness may be the cause of the somewhat
surprising shape of \textbf{Figure 1}, which shows a dip in status
completion among Alaskans around the time of the initiation of the PFD,
which is denoted by a vertical dotted line:

\begin{center}\includegraphics[width=0.8\linewidth]{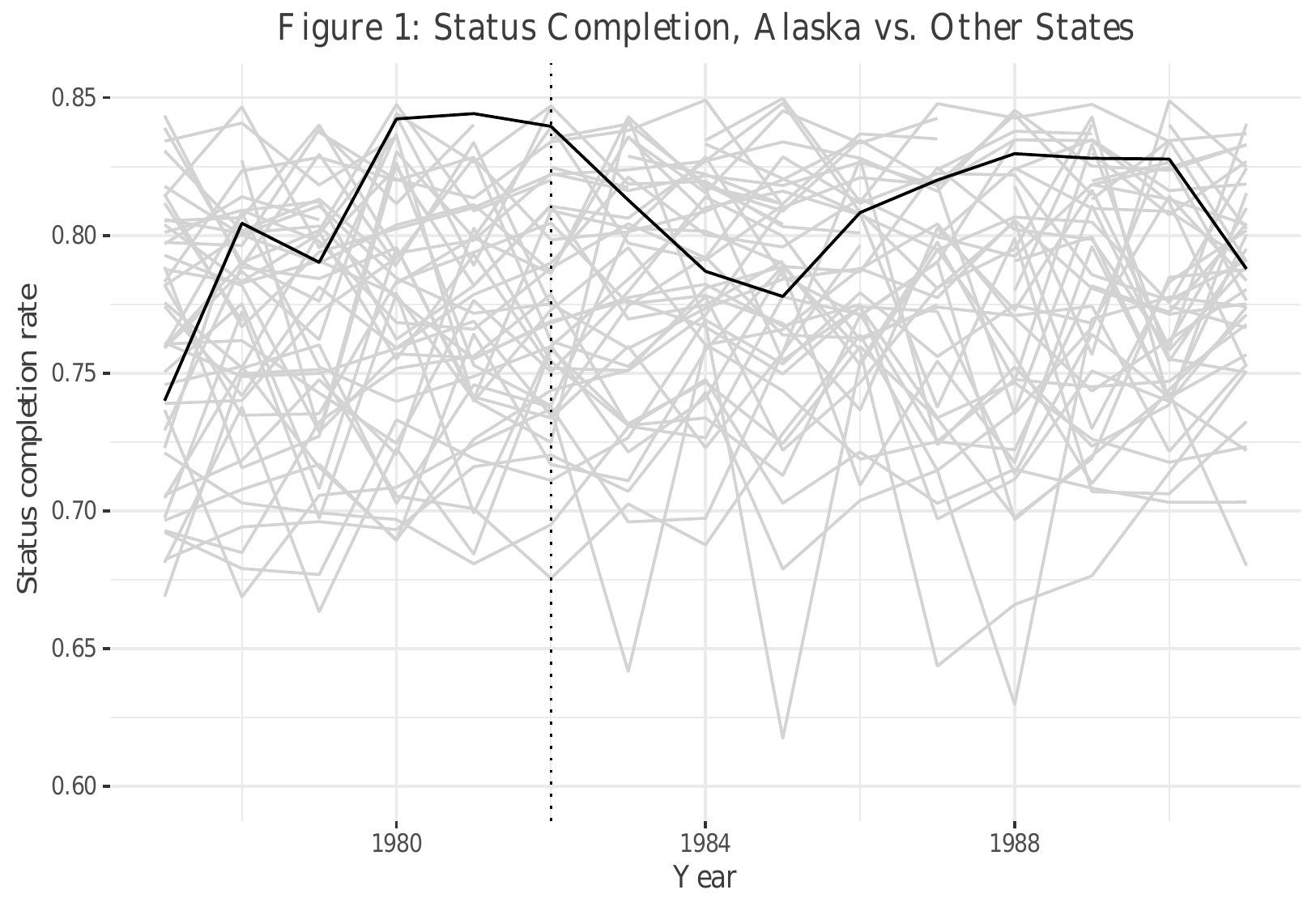} \end{center}

This plot also demonstrates the volatile nature of status completion
rates across states. Note the unusual increase in status completion
rates in the pretreatment period, e.g.~the first five years following
the completion of the Trans-Alaska Pipeline. This education spike can
perhaps be explained by the departure of pipeline workers in 1977
(Sandberg, 2013). As I will explain in more detail in the next section,
many of the laborers who moved to Alaska to work on the pipeline project
had relatively low levels of education; they came in very large numbers
and their gradual departure--as well as their aging out of the sample of
18- to 24-year-olds used to calculate status completion--may well have
resulted in an increase in that statistic.

This is just one of many analytical challenges that I will describe in
detail in the next section. Nevertheless, for the time period under
investigation this is the best available cross-state proxy measure for
high school completion, and it is the the statistic that I calculate for
each state using CPS data with appropriate weights and use in my
analysis below.

\subsection{Challenges and
Idiosyncracies}\label{challenges-and-idiosyncracies}

This research is complicated by two factors: the limitations of the
available data, and the somewhat unusual nature of the scenario under
investigation. As I detailed in the Introduction to this study, the
Permanent Fund Dividend came about as a consequence of the expected
completion of the Trans-Alaska Pipeline System (TAPS), which began
operating in 1977---perhaps not coincidentally the first year for which
disaggregated data is available for Alaska in the Current Population
Survey.

The introduction of oil revenues into the Alaskan economy was an
economic shock with far-reaching impacts, including, possibly, effects
on educational attainment that could plausibly swamp the influence of
the PFD itself in any state-level analysis. Yet the state is by no means
unique: at the time of the opening of the TAPS, several states produced
substantially more crude oil than Alaska; Texas alone produced more than
twice as much. Around 1977, Texas, Oklahoma, and New Mexico saw
comparable increases in crude oil production (in absolute terms) to that
experienced in Alaska. The value of Alaska's 1977 crude oil production
came out to roughly 0.7\% of the state's GDP in that year.

Nevertheless, TAPS was ``the most expensive privately financed
construction project in world history'' and therefore, given the small
size of the Alaskan economy, ``the largest localized demand shock in
postwar U.S. history'' (Carrington, 1996, p.~187). Thus, the large-scale
labor force and population changes wrought by the construction of the
pipeline and by the rapid growth of the petroleum industry in Alaska are
of considerable concern for the present analysis.

Construction on the TAPS began in 1974, at which time the Alyeska
Pipeline Service Company, which constructed and continues to operate the
pipeline, began to hire staff to work on the pipeline. Each summer of
the project, Alyeska and its subcontractors employed a total of roughly
50,000 individuals who fell, generally, into one of two groups. The
first group were primarily skilled welders and pipefitters, and
generally hailed from out-of-state (in particular from Texas and
Oklahoma). The second group consisted of unskilled laborers, and were
more likely to be Alaskans (Carrington, 1996). Together, these workers
would have constituted more than a tenth of Alaska's 1977 population.

Though the U.S. Census identifies respondents' birth states, it is
conducted only once every decade. The Current Population Survey does not
collect respondents' U.S. state of birth and only began to collect
respondents' one-year interstate migration data in 1982. It is therefore
difficult to obtain estimates of the proportion of pipeline workers who
remained in Alaska after the completion of the pipeline in 1977, the
rate at which those who left the state did so, and the amount that
stayed indefinitely.

This is a potential issue for the analysis I conduct below. Though
welding was and is considered skilled labor, CPS data shows that, in
1977, 57\% of welders had completed twelfth grade, in contrast to 67\%
of the U.S. population over 18 and 83\% of Alaskans over 18. The median
age of a U.S. welder in 1977 was 34 and 20\% were under 24. By 1982,
almost all of the migrant welders who were under 24 in 1977 would have
aged out of the sample of 18-to-24 year olds I use to calculate total
status completion. However, if some nontrivial portion of them remained
in Alaska after the completion of the pipeline, there remains the
concern that their presence could artificially suppress status
completion rates in the pretreatment period.

\section{Results}\label{results}

In this section, I compare educational outcomes in a synthetic Alaska
formed using a weighted combination of donor states (hereafter the
``control'') to those observed in the real Alaska (the ``treated
unit''). In \textbf{Table 1} I compare the treated unit to its control
using the averaged values of predictive covariates over the course of
the pre-treatment period, while \textbf{Table 2} contains the states
used to form the control and their accompanying weights.

\begin{longtable}[]{@{}lrrr@{}}
\caption{Pretreatment Predictor Values}\tabularnewline
\toprule
& Treated & Synthetic & Sample Mean\tabularnewline
\midrule
\endfirsthead
\toprule
& Treated & Synthetic & Sample Mean\tabularnewline
\midrule
\endhead
Log of Gross State Product & 23.209 & 23.042 & 24.134\tabularnewline
Unemployment rate & 9.720 & 6.476 & 6.290\tabularnewline
Percent of population over 24 with at least 4 years of college & 0.210 &
0.186 & 0.162\tabularnewline
State GDP per capita & 33888.804 & 14302.081 & 11107.228\tabularnewline
Status completion rate, 1977 & 0.740 & 0.757 & 0.776\tabularnewline
Status completion rate, 1981 & 0.844 & 0.841 & 0.782\tabularnewline
\bottomrule
\end{longtable}

\begin{longtable}[]{@{}rl@{}}
\caption{Nonzero State Weights}\tabularnewline
\toprule
Weight & State\tabularnewline
\midrule
\endfirsthead
\toprule
Weight & State\tabularnewline
\midrule
\endhead
0.169 & Connecticut\tabularnewline
0.543 & Delaware\tabularnewline
0.057 & Hawaii\tabularnewline
0.230 & Wyoming\tabularnewline
\bottomrule
\end{longtable}

\textbf{Figure 2} shows the actual paths of the treated unit and the
placebos used to form its synthetic control. \textbf{Figure 3} plots the
path of Alaska against its synthetic control, with a dotted vertical
lign at the launch of the PFD in 1982, while \textbf{Figure 4} plots the
gaps between the synthetic and the treated unit.

\begin{center}\includegraphics[width=0.8\linewidth]{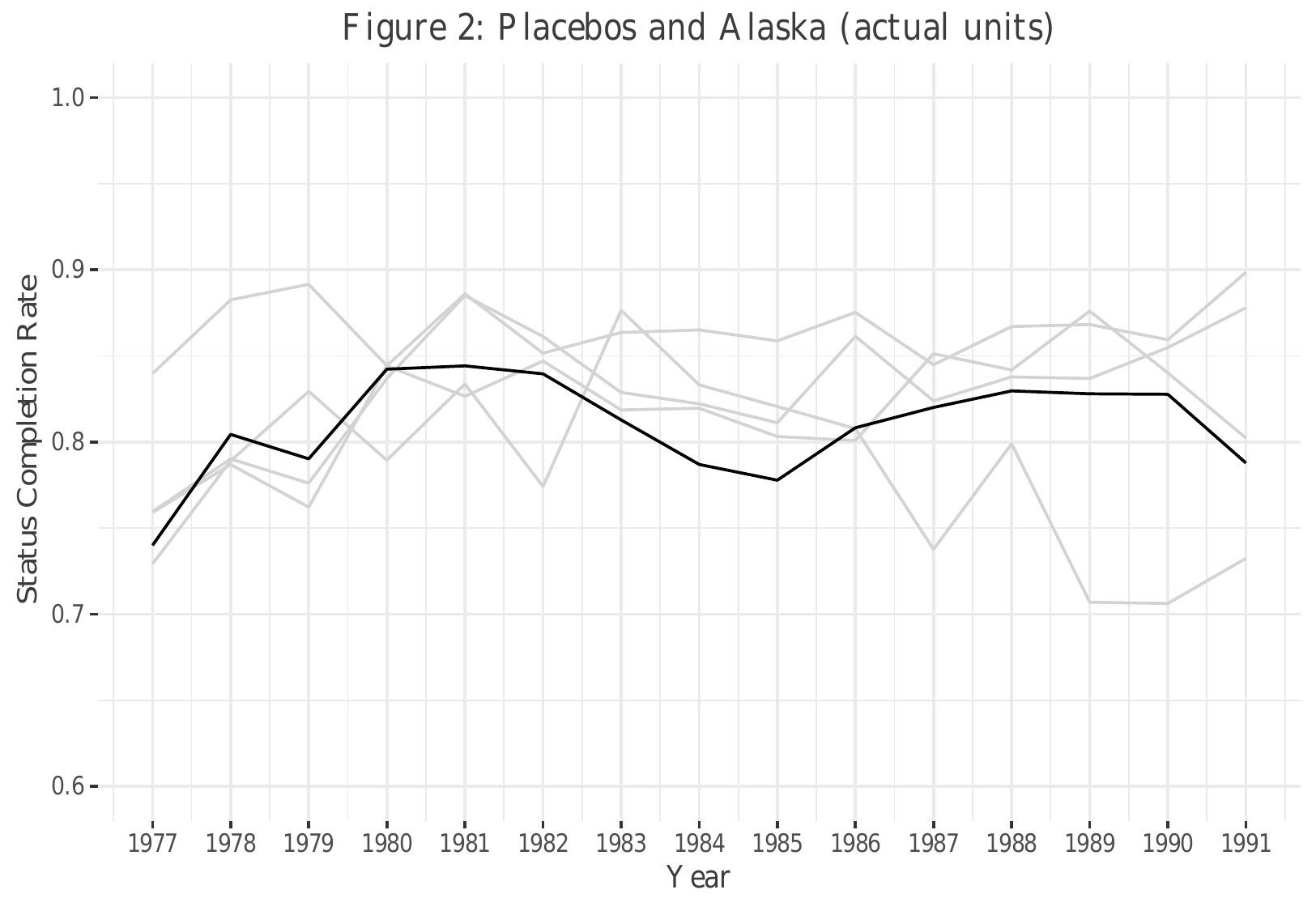} \includegraphics[width=0.8\linewidth]{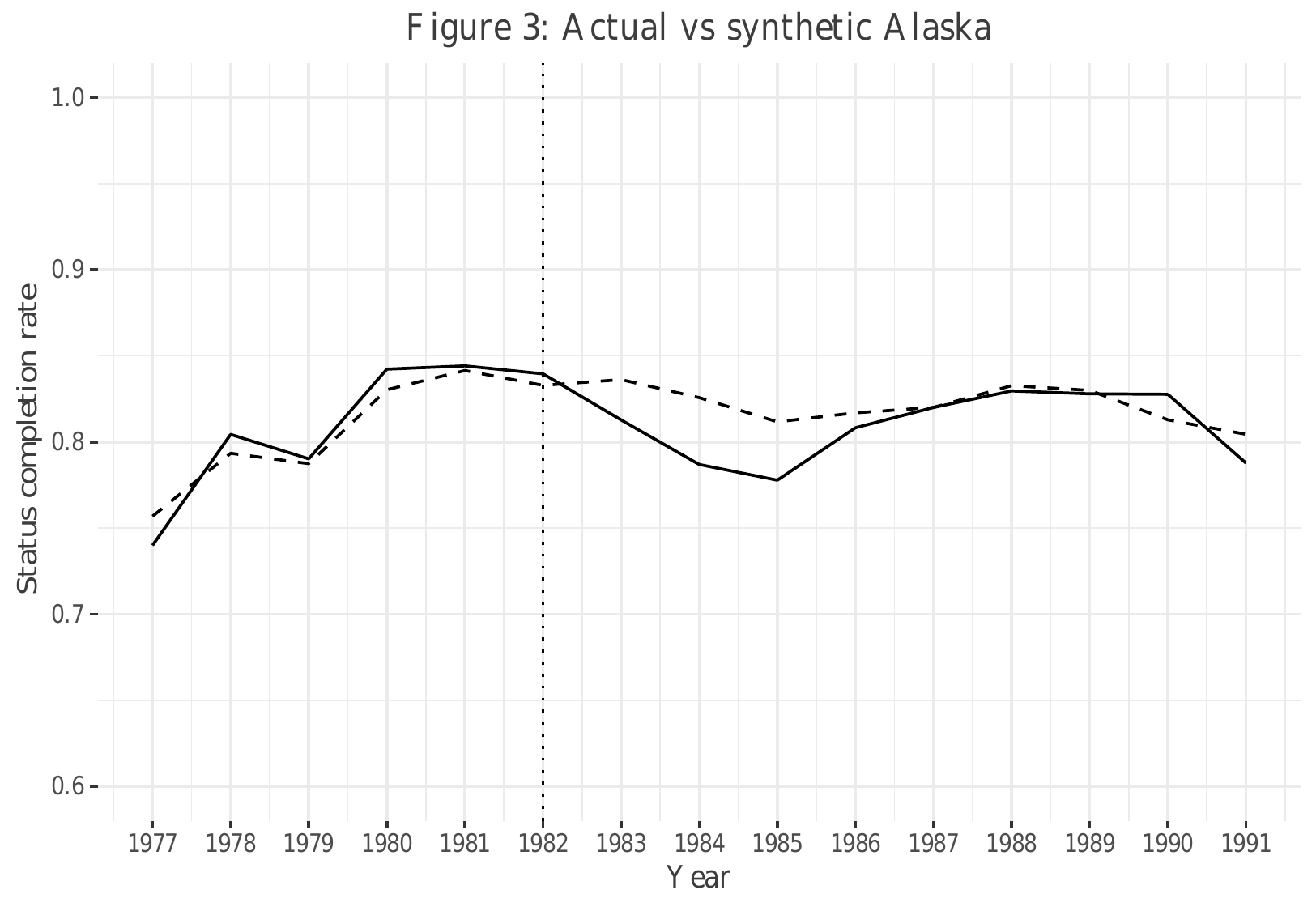} \includegraphics[width=0.8\linewidth]{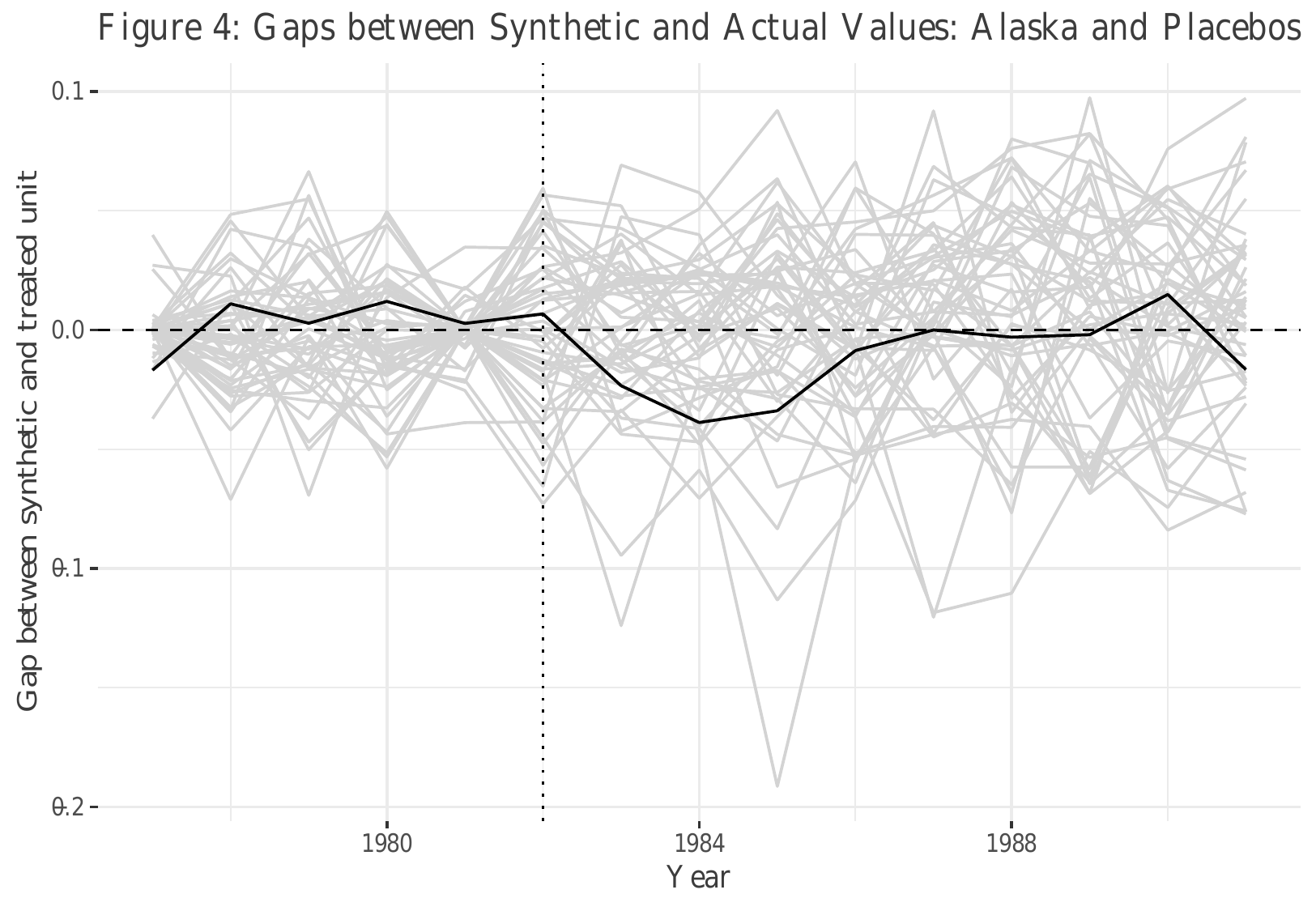} \end{center}

Visual inspection of \textbf{Figure 3} and \textbf{Figure 4} reveals no
easily discernible difference in outcome variable values between the
treated unit and its control. To quantify the degree to which this is
the case, I use the placebo method: for each actually untreated state in
the donor pool, I create a unique synthetic control model using the same
specifications as those for Alaska, determining the deviation between
the donor state and its control for the counterfactual situation in
which each of the donor states received the treatment. By taking as the
sampling distribution the small universe of possible synthetic-treated
gaps, I develop a ``p-value'' that suggests the likelihood of observing
a gap at least as large as the observed one (between Alaska and its
control) in event of a null effect.

In order to develop a usable p-value, I calculate the root mean squared
error (RMSE) for each placebo, and for the Alaska-control pair, using
the difference between synthetic and actual values for each year of the
post-treatment period under investigation (1982-1991).

With a placebo-based p-value of 0.95, I find what amounts to a null
effect: in placebo trials, RMSEs at least as large as the observed gap
between Alaska and its control are observed in 95\% of cases. The path
of educational attainment in Alaska between 1977 and 1991 therefore
closely tracks the experience of the real Alaska and does not
meaningfully depart with the launch of the Permanent Fund Dividend in
1982 or at any point thereafter.

As a further check, I report the ratio between post- and pre-treatment
RMSE for all possible synthetic controls, as suggested by Abadie,
Diamond \& Hainmueller (2015), in order to contextualize the Alaska
case. The principle at work here is that, under a non-null effect, that
paths of the posttreatment synthetic and treated unit will diverge,
resulting in a high ratio relative to the placebo regions.
\textbf{Figure 5} displays the RMSE ratios for Alaska and all placebos.

\begin{center}\includegraphics[width=0.8\linewidth]{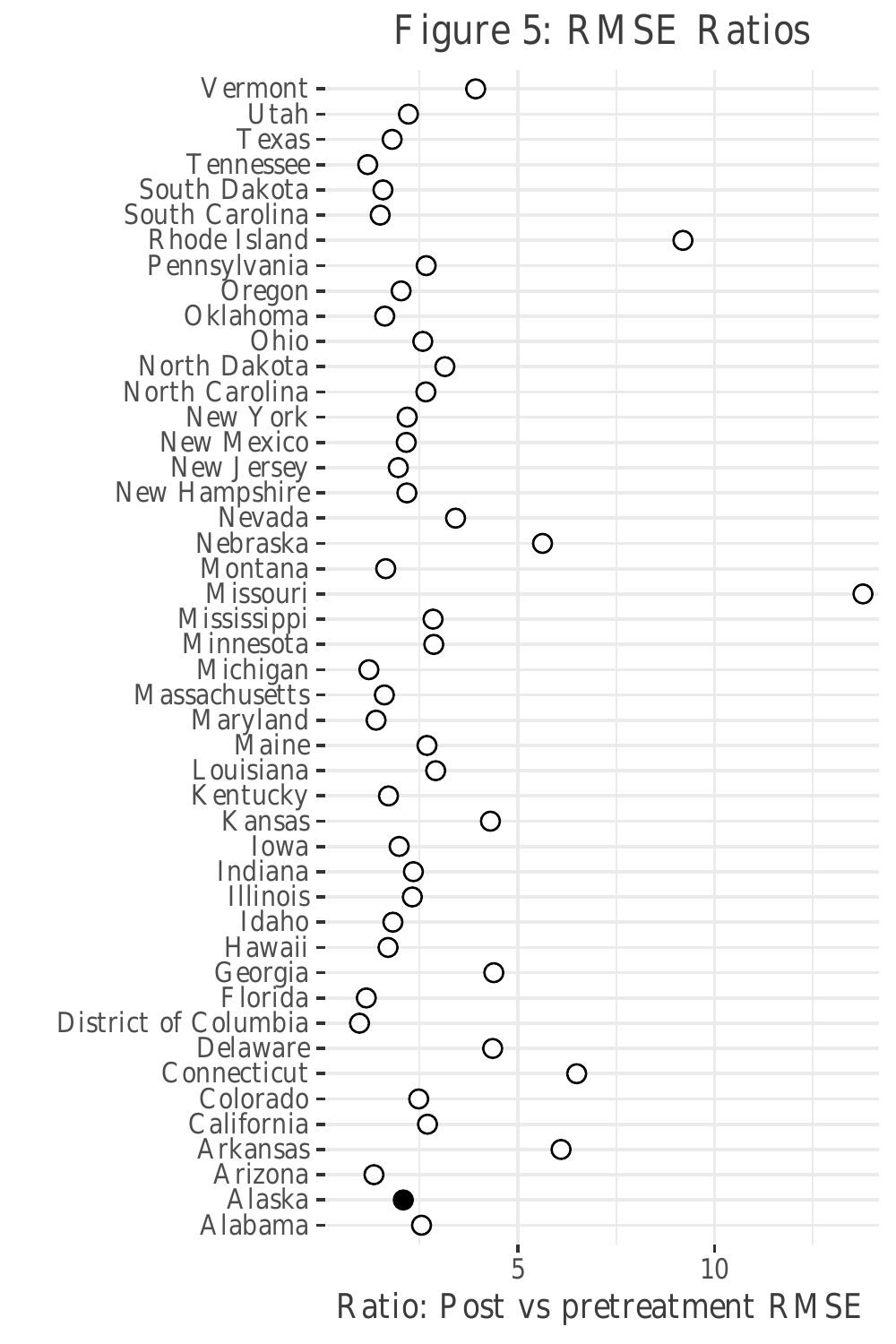} \end{center}

This plot demonstrates that the gap between Alaska and its synthetic
control unit does not meaningfully increase in the posttreatment period
as compared with a set of placebos fit to their controls using the same
set of covariates. In conjunction with the placebo tests and informal
p-value derived above, I find that the results of my analysis are not
suggestive of a positive effect of the initiation of the Alaska
Permanent Fund Dividend on high school status completion rates.

\subsection{Model Fit}\label{model-fit}

There is no widely agreed-upon rigorous criterion for assessing the fit
of a synthetic control model. In this particular case, the challenge is
complicated further by the relatively short (five-year) length of the
pretreatment period (an obstacle also faced by Jones and Marinescu
(2018)). In this section, I present several different measures of model
fit and present a rationale for why these metrics suggest that the
synthetic control model applied above is a good fit.

The first metric I explore is the pretreatment root mean squared error
(RMSE), or the square root of the sum of squared differences between the
synthetic control and the treated unit in the pretreatment period. In
the histogram below, I plot the distribution of the RMSE for Alaska and
for all placebo regions; the dotted vertical line in \textbf{Figure 6}
denotes the RMSE of the synthetic Alaska.

\begin{center}\includegraphics[width=0.8\linewidth]{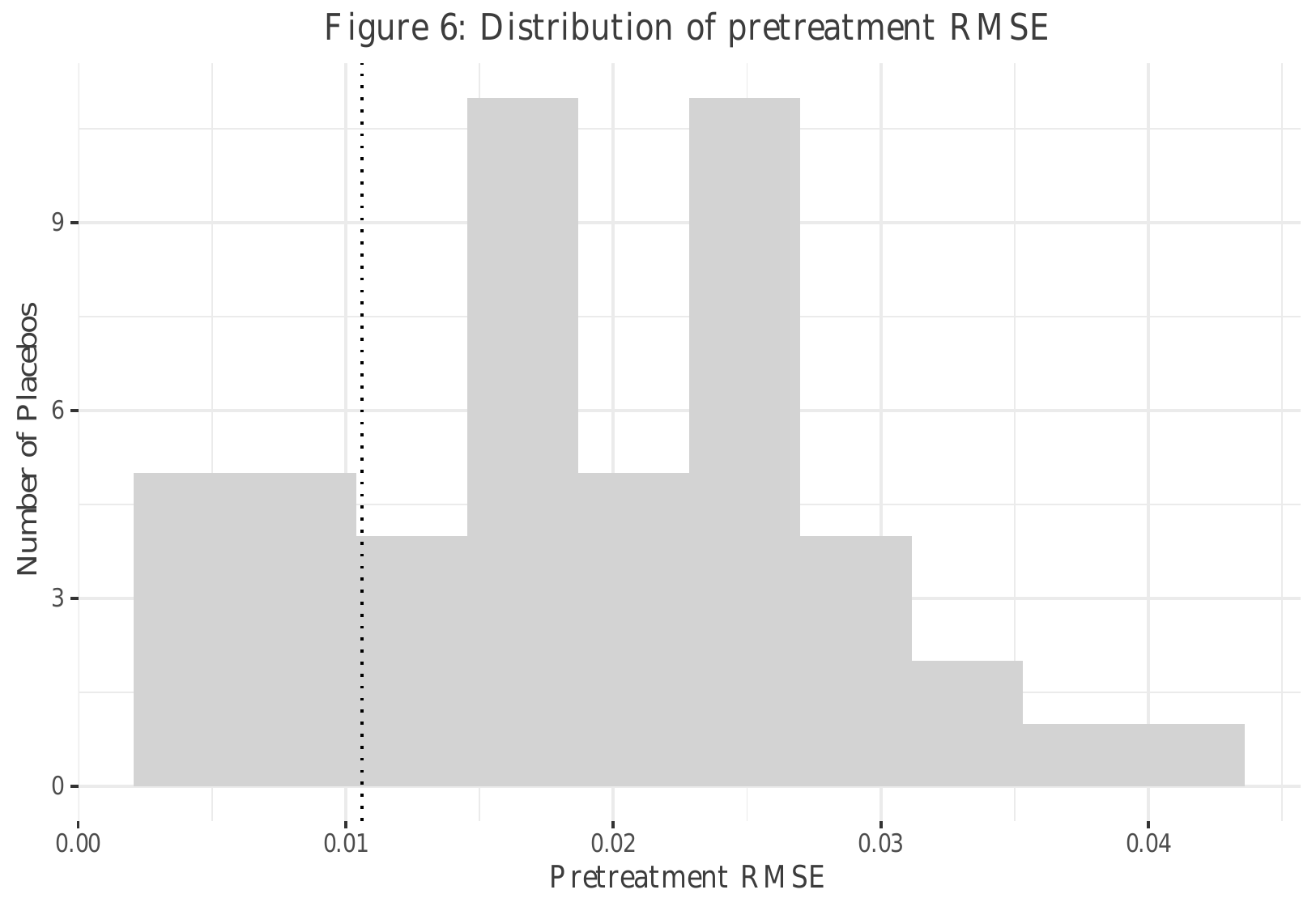} \end{center}

RMSEs for Alaska and its placebos range between 0.001 and 0.04, with a
mean of 0.02 and a value of 0.01 for Alaska itself. Thus, during the
pretreatment period for all placebos, the set of covariates selected for
the model produces synthetic control units within 0.1\% and 4\% of the
actual status completion rates for all regions, and within 1\% for
Alaska.

In order to complement the RMSE measurements above, I also include the
same set of statistics with respect to mean absolute error. Willmott \&
Matsuura (2005) argue that RMSE varies not only with the average error
magnitude but also with the square root of the number of errors and the
variability of the distribution of error magnitudes. I present MAE here
in order to preëmpt these concerns and for purposes of comparison with
RMSE.

\begin{center}\includegraphics[width=0.8\linewidth]{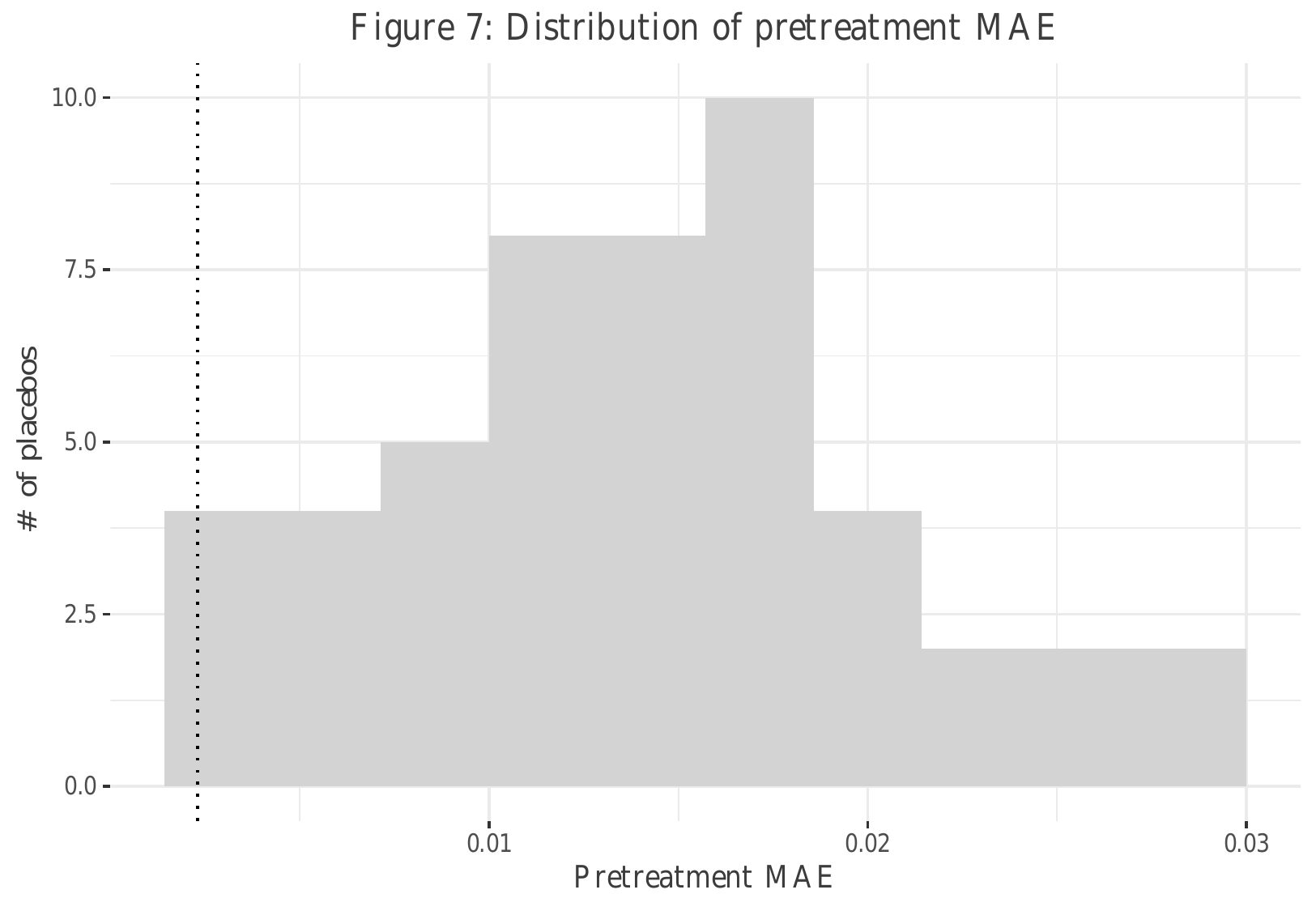} \end{center}

MAEs for Alaska and its placebos range between 0.003 and 0.028, with a
mean of 0.014; synthetic control units fall within 0\% and 2.8\% of the
actual status completion rates for all regions.

In order to offer a framework for interpreting these fit statistics, I
suggest considering RMSE within the context of the actual
period-to-period variation of the outcome metric for the treated unit in
the pretreatment period. Many practitioners of the synthetic control
method have tended to offer \emph{de facto} visual arguments for model
fit; Abadie, Diamond \& Hainmueller (2010), for example, offer
convincing plots depicting the apparently parallel pretreatment
trajectories of synthetic and treated units as an informal argument for
good model fit.

RMSE does not offer useful information about model fit in the absence of
a comparison with a competing model and that model's RMSE. However, the
relationship between this deviation and the actual variability of the
treated unit can be considered a good formal proxy for visually assessed
``fit.'' In the present context, for example, a synthetic control RMSE
of 1 percentage point would make for an impressive fit if the actual
treated unit had a standard deviation 5 percentage points, but would
represent an extremely poor fit if the unit were to tend to vary only
within about 0.1 percentage points.

This perspective offers a simple rule of thumb for considering whether a
synthetic control is a good fit for the treated unit: whether the RMSE
of the synthetic unit is smaller---preferably much smaller---than the
standard deviation of the treated unit in the pretreatment period. This
rule of thumb is well-suited to scenarios, such as the present one, in
which the available pretreatment period is short. For studies with more
extensive data, Li (2017) has proposed a subsampling method based on the
bootstrap that is suitable for use with a synthetic control.

The statistic I deploy below has been previously been suggested in the
statistical hydrology literature as a method for evaluating performance
of watershed models, which entail the simulation of monthly discharge
rates for a given waterway for the purposes of future flow forecasting
and attendant practical goals such as land management (Moriasi et al
2007). In this literature the figure is referred to as the
RMSE-observations standard deviation ratio, or RSR.

For clarity, I present this formally below: I calculate the statistic
\(RSR\) for a given state over \(T\) time periods as below, where
\(y_t\) represents the value of the variable of interest for the treated
unit at time \(t\) and \(\hat{y_t}\) is the value of the synthetic unit
at time \(t\).

\[ RSR = \frac{\sqrt{\frac{1}{T}\sum_{j = 0}^{T}{(\hat{y_t}  - y_t)^2}}}{\sqrt{\frac{1}{T-1}\sum_{t = 0}^{T}{(y_t  - \overline{y})^2}}} \]

The distribution of this statistic for the pretreatment period of the
synthetic Alaska is depicted below in Figure X, with a dotted vertical
line representing the value of this statistic for Alaska.

\begin{center}\includegraphics[width=0.8\linewidth]{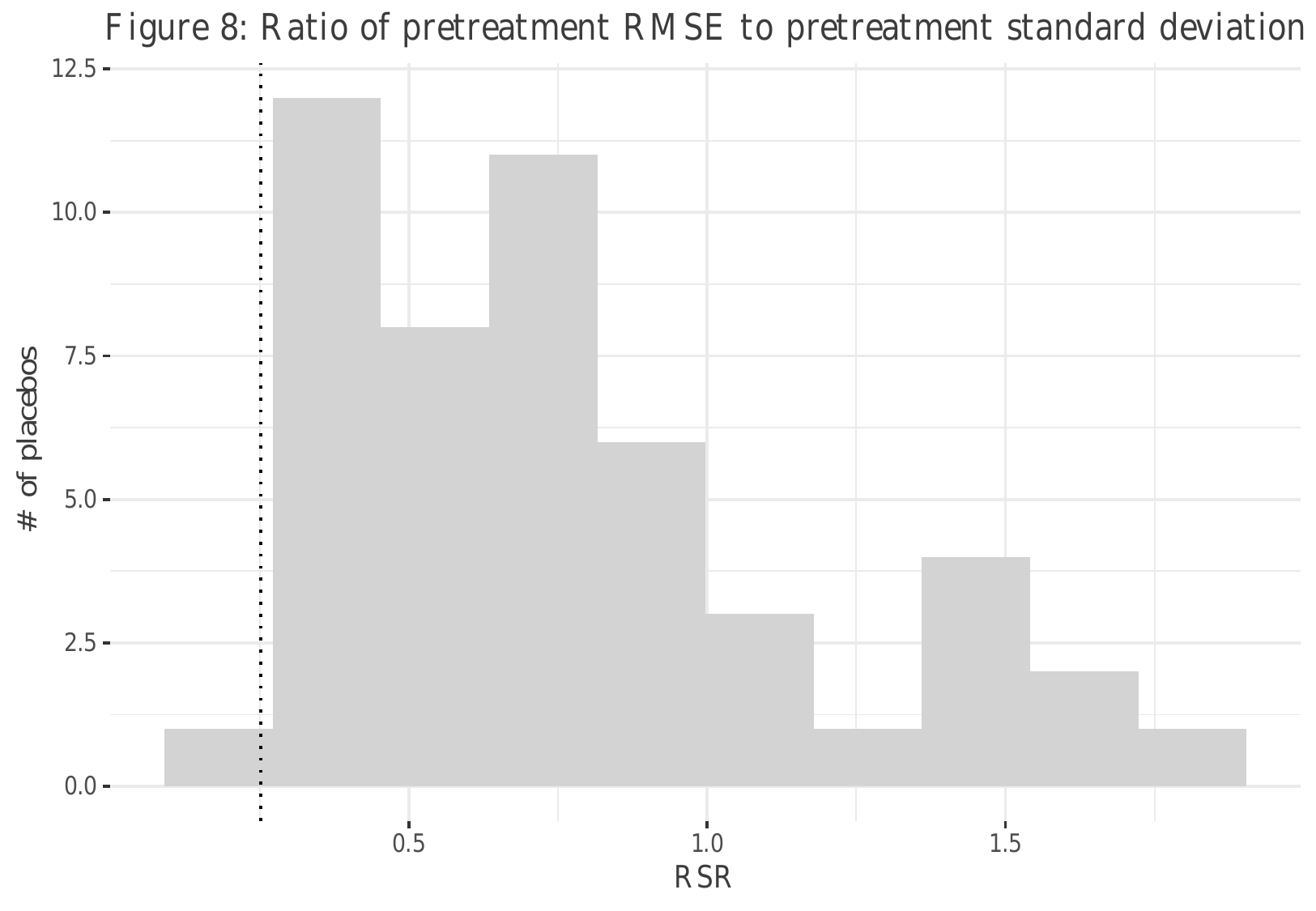} \end{center}

This histogram suggests that the collection of covariates used in this
synthetic control method is a good fit not just for Alaska, but for the
bulk of placebos as well. For nearly all regions, RSR values of less
than one indicate that the pretreatment variation between synthetic and
treated units is less, and often much less, than the natural variation
of the treated unit alone.

\subsection{Discussion}\label{discussion}

This study grows out of a large previous body of research on the social
welfare impacts of cash transfers. As I have detailed in my literature
review above, this research includes several distinct sub-areas. One
group of studies examines the effects of the Earned Income Tax Credit
(EITC), arguably the most successful antipoverty program in the history
of the United States. Researchers have found that even modest exposure
to the EITC in childhood can improve educational outcomes for low-income
children, and a host of other positive benefits attested in the academic
literature contrasts with the relatively limited amounts afforded to
low-income families by the disbursement of the EITC. This lends credence
to the idea that even modest cash transfers can have substantial effects
on affected populations.

The second group of studies includes the negative income tax trials
conducted in the United States in the 1970s. The results of many of
these studies indicated modest positive impacts of cash transfers on the
educational outcomes of low-income children, teenagers, and young
adults.

Lastly, there is the heterogeneous collection of various ``one-off''
studies on unconditional cash transfers. These include the recent and
substantial body of research on cash transfers in the developing world,
the Manitoba Basic Annual Income Experiment (``Mincome''), and programs
such as the annual payment disbursed by the Eastern Band of Cherokee
Indians in North Carolina.

This collection of disparate analyses, conducted over the course of
nearly half a century, provides strong prior evidence for the hypothesis
that relatively modest cash transfers---those, loosely speaking, between
\$1000 and \$5000---can positively impact educational attainment among
low-income youth, with an effect size on the order of about one extra
year of education.

The results of the synthetic control model above fail to provide
substantive support for this hypothesis. In this section, I address the
possibility of a true null effect, offer explanations for the null
finding in the hypothetical presence of a true non-null effect, and
offer methodological suggestions for future researchers.

\subsubsection{Considerations about the null
finding}\label{considerations-about-the-null-finding}

The simplest explanation for the null finding of my analysis is that the
synthetic control method that I use in this study is too blunt an
instrument to detect what could be a very small effect. With better
covariates or a superior dataset, it might be possible to develop a
tighter-fitting control unit, and therefore to discern a positive
effect. There is also space here to deploy other econometric methods: in
particular, it may be possible to use data from the American Community
Survey to understand the possible correlation between years of exposure
to the dividend and high school status completion.

Another possible explanation for the null finding of this analysis may
be that, if there are educational effects resulting from the
disbursement of the PFD, they are distributed heterogeneously across the
population. The research discussed in the previous section has focused
largely on the educational attainment of students from low income
families; it is plausible, if the distribution of the PFD has any effect
at all, that this effect is heterogeneous and acts only on low-income
students. Yet this explanation itself requires an explanation. Holding
other factors constant, individuals with higher household income are
generally more likely to be high school status completers than those
with low household income. Even if the disbursement of the Alaska
Permanent Fund Dividend displays a heterogeneous treatment effect on
educational attainment, with a trivial or null effect for middle- and
upper-income students, it would still be reasonable to expect overall
status completion rates to increase due to the increased completion
rates of low-income students.

A third possibility is that the size of the dividend is simply not large
enough to have a measurable effect on educational attainment. Yet this
presents a puzzle as well. Some literature on the EITC, for example,
suggests sizeable effects on high school completion associated with an
exposure to the credit, between the ages of 13 and 18, of approximately
\$1000 in 2018 dollars (Bastian \& Michelmore 2018). The families of
Alaskan children who were thirteen years old at the initiation of the
dividend in 1982 would have received the equivalent of more than \$6000
by the time those children turned 18 (Alaska Department of Revenue
2019), but, per the synthetic control model applied here, no positive
effect begins to reveals itself even as late as 1987.

A final possibility is that the prior evidence provided by research on
the NIT and the EITC is itself misleading. In my literature review, I
uncovered relatively few studies reporting null effects of the NIT or
the EITC. This could be an indication of the phenomenon known as the
file drawer effect or the file drawer problem (Rosenthal, 1979), in
which null results are not considered interesting and do not find their
way to publication. Those studies claiming education effects of the EITC
could be simply the non-null tip of a vast iceberg, the null bulk of
which is submerged. I am unconvinced by this possibility, given that
studies on EITC, NIT and MINCOME demonstrate broad concordance with the
respect to their reported effect sizes.

\subsubsection{Parameter of interest}\label{parameter-of-interest}

In previous sections, I have cited various challenges related to the
measurement of substantive state-level education statistics for the time
period in question. I feel that this is the primary obstacle to studying
this topic, and to my use of the synthetic control method in particular.
Status completion rate is the only available statistic for the period
under review that can be considered a good proxy for high school
completion, as well as an unusually blunt measure of that underlying
statistic in the case of Alaska. Yet status completion writ large is
not, perhaps, the most instructive statistic with respect to Alaskans'
educational attainment during the post-pipeline period: at the time of
the PFD's launch, Alaska had just wrapped up the largest privately
financed construction project in world history. Young men and a few
young women of differing levels of education were rapidly moving to and
from the state, and no effort was made by the Current Population Survey
to distinguish migrants from non-migrants (throughout the country) until
the greatest upheavals had already ended. For this reason, status
completion may not capture the underlying quantity of interest, which
might be summarized as ``the educational attainment of native-born
Alaskans''; sadly for my purposes, states of birth were not colleced by
the CPS during this time period, either.

The question of whether or not status completion is a suitable parameter
of interest therefore casts something of a pall over this research. Do
changes in the status completion rate reflect the impact of the PFD on
educational attainment, in whole or in part? The answer is surely
affirmative to some small extent, but numerous other factors affect
status completion---particularly so for the time and place I analyze
here. More importantly, the educational attainment of migrant laborers,
the vast majority of whom moved to Alaska after the age of 18, is not
really of interest, since none of the causal mechanisms proposed for the
apparent educational impacts of previous cash transfer programs could
have affected them. Nevertheless, there is no rigorous way to remove
these individuals from the data. For this reason, the question of
whether changes in status completion can measure effects on educational
attainment for those Alaskans who we could plausibly expect to be
affected by Permanent Fund Dividend payments remains an open one.

\section{Conclusion}\label{conclusion}

The disbursement of the Permanent Fund Dividend relies on Alaska's oil
revenues, the existence of which indexes a unique state of affairs that
makes comparison difficult. The launch of Alaska's sovereign wealth fund
and the initiation of the dividend in 1982 occurred in a period that
brought enormous upheaval to Alaska. With the start of construction on
the Trans-Alaska Pipeline System in 1974, workers from all over the
country swelled Alaska's population. At the same time, well-paid
pipeline jobs became available to native-born Alaskans and the state's
economy experienced a demand shock of rare magnitude. The Current
Population Survey only began disaggregating data for Alaska in 1977, and
only began to collect information about one-year interstate migration
patterns in 1981. At the time of the launch of the PFD, Alaska was a
unique region, and to some degree it is still today: it is
geographically isolated from the contiguous United States, sparsely
populated, and climactically inhospitable.

All of these factors represented challenges to my effort to study the
impact of the Permanent Fund Dividend on high school status completion
rates in the period following the start of payments in 1982. In this
paper, I fit a synthetic control to Alaska in the pretreatment period
1977-1981 and tracked this control in parallel with the treated unit in
order to identify possible divergences, which would be suggestive of a
treatment effect. I found no such effect, and verified the lack of a
notable divergence between treated and synthetic units by conducting
placebo tests.

Despite the methodological and data difficulties encountered over the
course of this study, I contend that my results are suggestive of a lack
of a substantive effect of the PFD on educational attainment in Alaska,
but do not suggest that there is no meaninful effect at all. As I
discuss in the Model Fit section above, the synthetic control closely
tracks the path of the actual Alaska during both the pretreatment and
post-treatment periods. The gap between Alaska and its synthetic control
is not notably higher than the gap between any placebo region and its
own synthetic during the post-treatment period, suggesting at the very
least that the treatment effect of the dividend does not exceed the
variation in the estimator.

However, as the root mean-squared error for Alaska's synthetic control
in the post-treatment period is approximately 2\%, it remains possible
that the PFD could have had a true effect on high school status
completion rates of less than 2\%. This would be a non-neglible effect
size, given the persistent challenges of ensuring that students graduate
high school.

Future researchers who hope to address this question or similar ones
would, perhaps, be better served by investing time and effort in
securing novel or unexploited data in order to conduct their analyses of
the effects of the PFD. In my own explorations of this matter, I feel I
have exhausted the limited ability of the Current Population Survey to
answer questions about this place and this time. My own abortive
experiments with the Panel Study of Income Dynamics and the American
Community Survey have led me to conclude that Alaska's
idiosyncracies---in particular, its status as a kind of statistical
backwater in the 1970s---make the use of large-scale survey data an
unpromising avenue of investigation.

Recently, as related by Forget (2011), social scientists have been able
to undertake an exhaustive reinvestigation of the Mincome experiment
using not the datasets originally produced by the project but, instead,
granular Canadian administrative data from the same place and time. I
recommend this as a path forward for future research into the welfare
effects of the Permanent Fund Dividend. For rigorous estimates of the
impact of the PFD, population-level statistics, as opposed to
survey-based estimates, may be necessary; it may also be necessary to
have access to official data in order to disentangle the population of
interest from Alaska's many comers and goers during the pipeline era.

For researchers interested in promoting or investigating the Alaska
Permanent Fund as a lens through which to view a possible future basic
income guarantee, I recommend not viewing the PFD as a panacea, if only
due to the many challenges of studying it. However, the PFD remains an
untapped well of data on the possible impacts of a society-wide
unconditional cash transfer program. It is the only program of its type
in the developed world, and research on its welfare impacts is
remarkably scarce. The same prior literature that led me to embark on
this path of research continues to provide strong evidence that cash
transfers are an effective means of improving outcomes, at least for
low-income households. What remains to be seen is whether means testing
and work requirements are cost-effective ways of disbursing such
programs, whether cash transfers can also be effective supports for
middle-income families, and whether society-wide transfer programs have
large-scale effects beyond their aggregate impact on individual
households. I hope future researchers will exploit the Alaska Permanent
Fund Dividend to help answer these questions.

\newpage

\section{Bibliography}\label{bibliography}

\setlength{\parindent}{-0.2in} \setlength{\leftskip}{0.2in}
\setlength{\parskip}{8pt} \noindent

Abadie, A., Diamond, A. \& Hainmueller, J. (2007). \emph{Synthetic
Control Methods for Comparative Case Studies: Estimating the Effects of
California's Tobacco Control Program} (NBER Working Paper No. 12831).
Retrieved from National Bureau of Economic Research website:
\url{https://www.nber.org/papers/w12831.pdf}

Abadie, A., Diamond, A. \& Hainmueller, J. (2010). Synthetic Control
Methods for Comparative Case Studies: Estimating the Effect of
California's Tobacco Control Program. \emph{Journal of the American
Statistical Association}, 105(490), 493-505.
\url{https://doi.org/10.1198/jasa.2009.ap08746}

Abadie, A., Diamond, A. \& Hainmueller, J. (2011). Synth: An R Package
for Synthetic Control Methods in Comparative Case Studies. \emph{Journal
of Statistical Software}, 42(13), 1-17.
\url{http://www.jstatsoft.org/v42/i13/}.

Abadie, A., Diamond, A., \& Hainmueller, J. (2015). Comparative Politics
and the Synthetic Control Method. \emph{American Journal of Political
Science}. 59(2), 495-510. \url{https://doi.org/10.1111/ajps.12116}

Abadie, A. \& Gardeazabal, J. (2003). The Economic Costs of Conflict: A
Case Study of the Basque Country. \emph{American Economic Review},
93(1), 113-132. \url{https://doi.org/10.1257/000282803321455188}

Alaska Department of Revenue. (2019). \emph{Summary of Dividend
Applications and Payments}. Retrieved from
\url{https://pfd.alaska.gov/Division-Info/Summary-of-Applications-and-Payments}

Alaska Permanent Fund Corporation. (2019). Home. Retrieved from
\url{https://apfc.org/}

Akee, R.K.Q., Copeland, W.E., Keeler, G., Angold, A., \& Costello, E.J.
(2010). Parents' Incomes and Children's Outcomes: A Quasi-Experiment
Using Transfer Payments from Casino Profits. \emph{American Economic
Journal: Applied Economics}, 2(1), 86-115.
\url{https://doi.org/10.1257/app.2.1.86}

Amarante, V., Manacorda, M., Miguel, E. \& Vigorito, A. (2011). \emph{Do
Cash Transfers Improve Birth Outcomes? Evidence from Matched Vital
Statistics, Social Security and Program Data} (NBER Working Paper No.
17690). Retrieved from National Bureau of Economic Research website:
\url{https://www.nber.org/papers/w17690.pdf}

Araujo, M.C., Bosch, M. \& Schady, N. (2016). \emph{Can Cash Transfers
Help Households Escape an Inter-Generational Poverty Trap?} (NBER
Working Paper No. 22670). Retrieved from National Bureau of Economic
Research website: \url{https://www.nber.org/papers/w22670.pdf}

Baird, S., Ferreira, F.H.G., Özler, B. \& Woolcock, M. (2014).
Conditional, unconditional and everything in between: a systematic
review of the effects of cash transfer programmes on schooling outcomes.
\emph{Journal of Development Effectiveness}, 6(1), 1-43.
\url{https://doi.org/10.1080/19439342.2014.890362}

Bastian, J. \& Michelmore, K., (2018). The Long-Term Impact of the
Earned Income Tax Credit on Children's Education and Employment
Outcomes. \emph{Journal of Labor Economics}, 36(4), 1127-1163.
\url{https://doi.org/10.1086/697477}

Benhassine, N., Devoto, F., Duflo, E., Dupas, P. \& Pouliquen, V.
(2013). \emph{Turning a Shove into a Nudge? A ``Labeled Cash Transfer''
for Education} (NBER Working Paper No. 19227). Retrieved from National
Bureau of Economic Research website:
\url{https://www.nber.org/papers/w19227.pdf}

Berman, M. (2018). Resource rents, universal basic income, and poverty
among Alaska's Indigenous peoples. \emph{World Development}, 106,
161--172. \url{https://doi.org/10.1016/j.worlddev.2018.01.014}

Bruenig, M. (2017). Social Wealth Fund for America. \emph{People's
Policy Project}. Retrieved from
\url{https://www.peoplespolicyproject.org/projects/social-wealth-fund/}

Card, D. (1990). The Impact of the Mariel Boatlift on the Miami Labor
Market. \emph{Industrial and Labor Relations Review}, 43(2), 245-257.
\url{https://doi.org/10.1177\%2F001979399004300205}

Card, D. \& Krueger, A. (1994). Minimum Wages and Employment: A Case
Study of the Fast-Food Industry in New Jersey and Pennsylvania.
\emph{The American Economic Review}, 84(4), 772-793. Retrieved from
\url{https://www.jstor.org/stable/2118030?seq=1\#metadata_info_tab_contents}

Carrington, W.J. (1996). The Alaskan Labor Market during the Pipeline
Era. \emph{Journal of Political Economy}, 104(1), 186-218. Retrieved
from
\url{https://www.jstor.org/stable/2138964?seq=1\#metadata_info_tab_contents}

Chetty, R., Saez, E., \& Rockoff, J. (2011). \emph{New Evidence on the
Long-Term Impacts of Tax Credits} (IRS Statistics of Income White
Paper). Retrieved from
\url{https://www.irs.gov/pub/irs-soi/11rpchettyfriedmanrockoff.pdf}

Council of Economic Advisers. (1998). \emph{Good News for Low Income
Families: Expansions in the Earned Income Tax Credit and the Minimum
Wage}. Retrieved from
\url{https://clintonwhitehouse4.archives.gov/media/pdf/eitc.pdf}

Current Population Survey. (2019). Retrieved from
\url{https://www.bls.gov/cps/?\#}

de Mel, S., McKenzie, D., Woodruff, C. (2007). \emph{Returns to Capital
in Microenterprises: Evidence from a Field Experiment} (Policy Research
Working Paper; No. 4230). Retrieved from World Bank website:
\url{https://openknowledge.worldbank.org/handle/10986/7124}

Doll, J.J., Eslami, Z., \& Walters, L. (2013). Understanding Why
Students Drop Out of High School, According to Their Own Reports: Are
They Pushed or Pulled, or Do They Fall Out? A Comparative Analysis of
Seven Nationally Representative Studies. \emph{SAGE Open}, 1(15).
\url{https://doi.org/10.1177/2158244013503834}

Eckholm, E. (2007, April 17). Tax Credit Seen as Helping More Parents.
\emph{New York Times}. Retrieved from
\url{https://www.nytimes.com/2007/04/17/us/17poor.html?_r=1\&pagewanted=print}

Edmonds, E.V. (2006). Child labor and schooling responses to anticipated
income in South Africa. \emph{Journal of Development Economics}, 81(2),
386-414. \url{https://doi.org/10.1016/j.jdeveco.2005.05.001}

Finnish Ministry of Social Affairs and Health. (2019). \emph{The basic
income experiment 2018-2018 in Finland: Preliminary results}. Retrieved
from
\url{http://julkaisut.valtioneuvosto.fi/bitstream/handle/10024/161361/Report_The\%20Basic\%20Income\%20Experiment\%2020172018\%20in\%20Finland.pdf}

Flood, S., King, M., Rodgers, R., Ruggles, S., \& Warren, J.R. (2018).
\emph{Integrated Public Use Microdata Series, Current Population Survey:
Version 6.0} {[}dataset{]}. \url{https://doi.org/10.18128/D030.V6.0}

Forget, E.L. (2011). The Town with No Poverty: The Health Effects of a
Canadian Guaranteed Annual Income Field Experiment. \emph{Canadian
Public Policy}, 37(3), 283-305.
\url{https://doi.org/10.3138/cpp.37.3.283}

Frank, R. (2006, Nov 23). The Other Milton Friedman: A Conservative With
a Social Welfare Program. \emph{New York Times}. Retrieved from
\url{https://www.nytimes.com/2006/11/23/business/23scene.html}

Friedman, M. (2002). \emph{Capitalism and Freedom: Fortieth Anniversary
Edition}. Chicago, IL: University of Chicago Press.

Friedman, M. (1987). The Case for the Negative Income Tax (a view from
the right). In Leube, K. (Ed.), \emph{The Essence of Friedman}
(pp.~57--68). Retrieved from
\url{https://miltonfriedman.hoover.org/friedman_images/Collections/2016c21/1966decnegativeYT.pdf}

GiveDirectly. (2017). Basic Income. Retrieved from
\url{https://www.givedirectly.org/basic-income}.

Goodman, P.S. (2018, May 30). Free Cash to Fight Income Inequality?
California City Is First in U.S. to Try. \emph{New York Times}.
Retrieved from
\url{https://www.nytimes.com/2018/05/30/business/stockton-basic-income.html}

Hanushek, E. (2003, September). Non-Labor-Supply Responses to the Income
Maintenance Experiment. In Munnell, A.H. (Ed.), \emph{Lessons from the
Income Maintenance Experiments}. Lessons from the Income Maintenance
Experiments: Conference Series No. 30., Melvin Village, New Hampshire
(pp.~106-121). Retrieved from
\url{https://www.bostonfed.org/-/media/Documents/conference/30/conf30.pdf?la=en}

Hotz, V.J. \& Scholz, J.K. (2003). The Earned Income Tax Credit. In
Moffitt, R.A. (Ed.), \emph{Means-Tested Transfer Programs in the United
States} (pp.~141-197). Retrieved from
\url{https://www.nber.org/chapters/c10256.pdf}

Hoynes, H. (2014, Summer). A Revolution in Poverty Policy: The Earned
Income Tax Credit and the Well-Being of American Families.
\emph{Pathways}. Retrieved from
\url{https://inequality.stanford.edu/sites/default/files/media/_media/pdf/pathways/summer_2014/Pathways_Summer_2014.pdf}

Hoynes, H., Miller, D. \& Simon, D. (2017). Income, the Earned Income
Tax Credit, and Infant Health. \emph{American Economic Journal: Economic
Policy}, 7(1), 172-211.\url{http://dx.doi.org/10.1257/pol.20120179}

Hsieh, C. (2003). Do Consumers React to Anticipated Income Changes?
Evidence from the Alaska Permanent Fund. \emph{American Economic
Review}, 93(1), 397-405.
\url{https://doi.org/10.1257/000282803321455377}

Jones, D. \& Marinescu, I. (2018). \emph{The Labor Market Impacts of
Universal and Permanent Cash Transfers: Evidence from the Alaska
Permanent Fund} (NBER Working Paper No. 24312). Retrieved from National
Bureau of Economic Research website:
\url{https://home.uchicago.edu/~j1s/Jones_Alaska.pdf}

Jordan, M.P. and Grossmann, M. (2017). \emph{The Correlates of State
Policy Project v.2.1}. Retrieved from
\url{http://ippsr.msu.edu/public-policy/correlates-state-policy}

Kansaneläkelaitos. (2017). Basic Income Experiment 2017-2018. Retrieved
from \url{https://www.kela.fi/web/en/basic-income-experiment-2017-2018}

Kaufman, P., Alt, M.N. \& Chapman, C.D. National Center for Education
Statistics, U.S. Department of Education. (2004). \emph{Dropout Rates in
the United States: 2001}. Retrieved from
\url{https://nces.ed.gov/pubs2005/2005046.pdf}

Kaul, A., Klößner, S., Pfeifer, G., \& Schieler, M. (2015). Synthetic
control methods: Never use all pre-intervention outcomes together with
covariates. Retrieved from
\url{http://www.oekonometrie.uni-saarland.de/papers/SCM_Predictors.pdf}

Li, K.T. (2017). \emph{Estimating Average Treatment Effects Using a
Modified Synthetic Control Method: Theory and Applications}. Retrieved
from
\url{https://faculty.wharton.upenn.edu/wp-content/uploads/2017/06/6_Kathleen-T-Li-Job-Market-Paper-Modified-Synthetic-Control-Method-1.pdf}

Manoli, D. \& Turner, N. (2018). Cash-on-Hand and College Enrollment:
Evidence from Population Tax Data and the Earned Income Tax Credit.
\emph{American Economic Journal: Economic Policy}, 10(2), 242-271.
\url{https://doi.org/10.1257/pol.20160298}

Marr, C., Huang, C., Sherman, A. \& DeBot, B. (2015). EITC and Child Tax
Credit Promote Work, Reduce Poverty, and Support Children's Development,
Research Finds. \emph{Center on Budget and Policy Priorities}. Retrieved
from
\url{https://www.cbpp.org/sites/default/files/atoms/files/6-26-12tax.pdf}.

Maynard, R.A. \& Murnane, R.J. (1979). The Effects of a Negative Income
Tax on School Performance: Results of an Experiment. \emph{The Journal
of Human Resources}, 14(4), 463-476.
\url{https://doi.org/10.2307/145317}

McDowell Group. (2017, May). \emph{The Role of the Oil and Gas Industry
in Alaska's Economy}. Retrieved from
\url{https://www.aoga.org/sites/default/files/final_mcdowell_group_aoga_report_8.16.17.pdf}

Moffitt, R.A. (2003). \emph{The Negative Income Tax and the Evolution of
U.S. Welfare Policy} (NBER Working Paper No. 9751). Retrieved from
National Bureau of Economic Research website:
\url{https://www.nber.org/papers/w9751.pdf}

Moriasi, D.N., Arnold, J.G., Van Liew, M.W., Bingner, R.L., Harmel,
R.D., \& Veith, T.L. (2007). \emph{Transactions of the American Society
of Agricultural and Biological Engineers}, 50(3), 885-900. Retrieved
from
\url{https://pubag.nal.usda.gov/pubag/downloadPDF.xhtml?id=9298\&content=PDF}

National Center for Education Statistics. (2016). \emph{Trends in High
School Dropout and Completion Rates in the United States}.
\url{https://nces.ed.gov/programs/dropout/}

National Center for Education Statistics. (2018). \emph{The Condition of
Education}. \url{https://nces.ed.gov/programs/coe/indicator_coi.asp}

O'Brien, J. P. \& Olson, D.O. (1990). The Alaska Permanent Fund and
Dividend Distribution Program. \emph{Public Finance Quarterly}, 18(2),
139-56. \url{https://doi.org/10.1177\%2F109114219001800201}

Owusu-Addo, E., Renzaho, A.M.N., \& Smith, B.J. (2018). The impact of
cash transfers on social determinants of health and health inequalities
in sub-Saharan Africa: a systematic review. \emph{Health Policy and
Planning}, 33(5), 675--696. \url{https://doi.org/10.1093/heapol/czy020}.

Pega, F., Liu, S., Walter, S., Pabayo, R,. Saith, R. \& Lhachimi, S.K.
(2017). Unconditional cash transfers for reducing poverty and
vulnerabilities: effect on use of health services and health outcomes in
low- and middle-income countries. \emph{Cochrane Database of Systematic
Reviews}, 2017(11), 1-180.
\url{https://doi.org/10.1002/14651858.CD011135.pub2}

Rosenthal, R. (1979). The file drawer problem and tolerance for null
results. \emph{Psychological Bulletin}, 86(3), 638-641.
\url{http://dx.doi.org/10.1037/0033-2909.86.3.638}

Sandberg, E. Alaska Department of Labor and Workforce Development.
(2013, April). \emph{A History of Alaska Population Settlement}.
Retrieved from
\url{http://live.laborstats.alaska.gov/pop/estimates/pub/pophistory.pdf}

Sen, Amartya. (1989). Development as capability expansion.~\emph{Journal
of Development Planning}.~19~(1): 41--58.

U.S. Census Bureau (2011). \emph{2001-2005 American Community Survey}.
Retrieved from \url{https://usa.ipums.org}

U.S. Energy Information Administration. (2018). \emph{U.S. monthly crude
oil production exceeds 11 million barrels per day in August}. Retrieved
from \url{https://www.eia.gov/todayinenergy/detail.php?id=37416}

Venti, S.F. (1984). The Effects of Income Maintenance on Work,
Schooling, and Non-Market Activities of Youth. \emph{The Review of
Economics and Statistics}, 66(1), 16-25.
\url{http://doi.org/10.2307/1924691}

Willmott, C.J. \& Matsuura, K. (2005). Advantages of the mean absolute
error (MAE) over the root mean square error (RMSE) in assessing average
model performance. \emph{Climate Research}, 2005(30), 79-82. Retrieved
from \url{https://www.int-res.com/articles/cr2005/30/c030p079.pdf}

YC Research. (2017). The first study of basic income in the United
States. Retrieved from \url{https://basicincome.ycr.org/}

Zwolinski, M. (2014). The Pragmatic Libertarian Case for a Basic Income
Guarantee. \emph{Cato Unbound}. Retrieved from
\url{https://www.cato-unbound.org/2014/08/04/matt-zwolinski/pragmatic-libertarian-case-basic-income-guarantee}

\end{document}